\documentclass[journal=ancac3,manuscript=article]{achemso}

\usepackage[version=3]{mhchem} 
\usepackage{xcolor}
\usepackage[normalem]{ulem}

\newcommand{\added}[1]{{\color{teal}#1}}
\newcommand{\deled}[1]{{\color{red}\sout{#1}}}
\renewcommand{\added}[1]{{#1}}
\renewcommand{\deled}[1]{{}}
\newcommand{\mycite}[2]{\cite[][#2]{#1}}
\newcommand{\myeq}[2]{\cite[Ref.][Eq. (#2)]{#1}}
\renewcommand{\mycite}[2]{\cite{#1}}
\renewcommand{\myeq}[2]{\cite{#1}}

\author{Christian Bunker}
\affiliation[UF]
{Department of Physics, Center for Molecular Magnetic Quantum Materials, and Quantum Theory Project, University of Florida, Gainesville, Florida 32611}
\author{Silas Hoffman}
\affiliation{Laboratory for Physical Sciences,  College Park, Maryland 20740}
\author{Shuanglong Liu}
\affiliation
{Department of Physics, Northeastern University, Boston, Massachusetts 02115}
\author{Xiao-Guang Zhang}
\affiliation[UF]
{Department of Physics, Center for Molecular Magnetic Quantum Materials, and Quantum Theory Project, University of Florida, Gainesville, Florida 32611}
\author{Hai-Ping Cheng}
\affiliation
{Department of Physics, Northeastern University, Boston, Massachusetts 02115}
\email{ha.cheng@northeastern.edu}

\title{Using near-flat-band electrons for read-out of molecular spin qubit entangled states}

\abbreviations{}

\begin{document}





\begin{abstract}

While molecular spin qubits (MSQs) are a promising platform for quantum computing, read-out has been largely limited to electron paramagnetic resonance which is often slow and requires a global system drive. Moreover, because one prerequisite for the Elzerman and Pauli spin blockade readout mechanisms typical of semiconductor spin qubits is tunneling of electrons between sites, these read-out modalities are unavailable in MSQs. Here, we theoretically demonstrate electrical read-out of entangled MSQs \textit{via} driven many-electron spin unpolarized currents.
In particular, using a time-dependent density matrix renormalization
group approach we simulate a maximally entangled MSQ pair between two electronic leads. Driving itinerant electrons between the two leads, we find that the conductance is greater when the MSQs are in the entangled singlet state as compared to the entangled triplet state. This contrast in conductance is enhanced when the electronic density of states at the Fermi energy is large and for narrow bandwidth. Our results are readily applicable to molecules \added{supramolecularly functionalizing}\deled{deposited on} semiconductors with relatively flat bands such as single-wall carbon nanotubes under a magnetic field.

\end{abstract}

\section{Introduction}



\newcommand{\inref}[1]{Ref.~\citenum{#1}}

Itinerant electrons coupled to
quantum information \cite{nielsen, div_qis} stored in the spin degree of freedom of a localized particle such as a semiconductor spin qubit \cite{div_loss, ladd} or a molecular spin qubit (MSQ) \cite{sessoli, hill, loss_leuen} offer a promising avenue towards functional quantum information science devices \cite{sutton}.
Electrons are particularly well suited to generating and probing entanglement between pairs of spin qubits \cite{costa, ciccarello2, yuasa, yuasa2, bunker, ciccarello}, a key task for quantum information science.
For example, maximally entangling two qubits using an electron as a measurement ancilla has been proposed \cite{costa, ciccarello2} and demonstrated 
\cite{bernien_measurement}.
Further, using itinerant electrons to electrically probe and read-out qubit states enables compact, scalable qubits \cite{coronado_swap, wernsdorfer_msq,  sutton} based largely on architectures that have already been optimized by the semiconductor industry \cite{dzurak_arch_proposal, dzurak_electrical, adv_semi_mfing}.

\newcommand{\supramolecular}{\cite{wernsdorfer2, wernsdorfer_graphene, wernsdorfer_gmr, liu_Mn12}\,}
\newcommand{\noncovalent}{\cite{chemical_reviews, jacs_2009, jmaterchem_2011,  acsnano_2010, jacs_2010, wernsdorfer_graphene}\,}
\newcommand{\molecularjunctions}{\cite{wernsdorfer_msq, wernsdorfer_nuc, wernsdorfer_nuc2, wernsdorfer_algo}\,}
\newcommand{\tddmrgtransport}{\cite{feiguin, shuai, burrello, feiguin_wilsonchain, feiguin_conductance}\,}
\newcommand{\tddmrgenergy}{\cite{shuai, burrello}\,}

For semiconductor spin qubits, spin-to-charge conversion \cite{ladd, petta} is a trusted electrical read-out protocol requiring tunneling of the qubit electrons.
Unlike in semiconductors, MSQs lack convenient tunneling barrier control,
so facilitating tunneling is difficult \cite{wernsdorfer_msq}.
Alternatively, a measurement of the \emph{conductance} past or through the MSQs avoids tunneling \cite{wernsdorfer_msq}.
Conductance measurements have been experimentally demonstrated to read-out the quantum state of \emph{single} spins \cite{wernsdorfer_msq},
but extending conductance-based read-out to multi-spin molecular qubit encodings such as singlet-triplet qubits
remains to be demonstrated.

Singlet-triplet ($ST_0$) qubits \cite{ladd} embed one logical qubit
within the maximally entangled two-spin states
\begin{flalign}
    |\psi_\text{qubit}\rangle=(|\uparrow_1 \downarrow_2 \rangle + e^{i\phi_\text{ent}}|\downarrow_1 \uparrow_2 \rangle )/\sqrt{2}
    \label{eq:cicc_state_alt}
\end{flalign}
where $|\psi_\text{qubit}\rangle = |S\rangle$ ($\phi_\text{ent}=\pi$) and $|\psi_\text{qubit}\rangle = |T_0\rangle$ ($\phi_\text{ent}=0$) 
define the logical states.
This qubit encoding benefits from fast state preparation and read-out as well as minimal decoherence under
global magnetic field fluctuations \cite{loss_tarucha}.
Electrical read-out of semiconductor $ST_0$ qubits relies on Pauli spin blockade \cite{petta, loss_tarucha, nichol_prapp, ladd_PSB}. 
However, since Pauli spin blockade requires the tunneling control typically lacking in MSQ architectures, the ability to measure conductance for electrical read-out of molecular $ST_0$ qubits
in addition to single spins
remains a desirable objective.

MSQ read-out \textit{via} correlation between the conductance and $|\psi_\text{qubit}\rangle$ would require interaction between entangled MSQ states and itinerant electrons,
\deled{This interaction can be realized in supramolecular devices \supramolecular where itinerant electrons flow through a nanowire and the MSQ ligands electronically couple them to the nanowire }\added{such as those arising in  supramolecular devices \supramolecular comprising semiconducting nanowires, for example single-wall carbon nanotubes or graphene nanoribbons, functionalized by MSQs.}\deled{Supramolecular devices benefit from the tunable electronic properties of $sp^2$ carbon materials such as single wall carbon nanotubes (SWCNTs) \cite{roche_revmod} and leverage the inherent scalability and chemical customization of MSQs \cite{hill, zadrozny, sessoli_van, coronado_swap}}
\deled{ (Fig.~\ref{fig:setup_emul}), many itinerant electrons traverse the nanowire without hopping onto the MSQs. Instead, a virtual hopping process yields an effective $sd$ exchange between the itinerant electrons and MSQs.}
%
\added{These systems feature MSQ spins exerting effective torque on the itinerant electron spins known as $sd$ exchange \cite{wernsdorfer2}, similar}
\deled{$sd$ exchange is often compared}\cite{schrieffer, switzer} to the Kondo interaction \cite{kondo, hewson} between itinerant electrons and magnetic impurities in metals,
\added{and}\deled{and has been} measured in the 0.1-1 meV range in supramolecular devices \cite{wernsdorfer2, liu_Mn12}.
\added{We describe specific chemical realizations of $sd$ exchange for MSQs deposited on semiconductor nanowires in the discussion.}
%
%
\inref{ciccarello} was the first to show that $sd$ exchange \added{links the}\deled{correlates}
conductance \added{to the}\deled{with} $ST_0$ qubit states (\ref{eq:cicc_state_alt}),
showing that for $\phi_\text{ent}=\pi$, constructive interference makes the conductance essentially perfect, while $\phi_\text{ent}=0$ impedes the conductance \cite{ciccarello, ciccarello_interferometer}.
\added{\inref{ciccarello} further demonstrated that this conductance disparity grew with increasing density of states in the semiconductor nanowire.}
%
%
The conductance dependence on $|\psi_\text{qubit}\rangle$, modulated by $\phi_\text{ent}$ \cite{ciccarello, ciccarello_interferometer, tulapurkar_cicc}, is a quantum analogue of a classical spin valve in which conductance depends on the relative polarization between two ferromagnetic leads \cite{parkin_GMR, grunberg_GMR, fert_review, spintronics_review, baumgartner}.
Consequently, we refer to the conductance dependence discovered in \inref{ciccarello}
as the \emph{quantum spin-valve effect}.
Just as the classical spin valve effect can be leveraged for read-out of classical bits, the quantum spin-valve effect enables read-out of $ST_0$ qubits. 

Despite the exciting applications of quantum spin valves to reading-out molecular $ST_0$ qubits, understanding of these systems remains immature.
\added{Supramolecular devices \supramolecular have experimentally demonstrated classical spin valve physics but have not yet shown conductance measurements sensitive to inherently quantum states such as the entangled $|S\rangle$ and $|T_0\rangle$ states.}\deled{Experimentally, supramolecular devices have demonstrated classical spin valve physics \cite{wernsdorfer2, liu_Mn12} but not quantum spin valves.}
On the theory side,  \inref{ciccarello}
\added{used a single-particle wavefunction matching method to simulate the conductance of a single electron only,
which cannot speak to the viability of the quantum spin valve effect and its application to read-out in a realistic many-electron semiconductor nanowire.}\deled{ demonstrated only single-electron conductance with a single-particle wavefunction matching method \cite{ciccarello, menezes}, failing to confirm the viability of the quantum spin valve effect and its application to read-out in a realistic many-electron device.}
%
%
To build on this work, we use many-body transport methods
to simulate entangled MSQs interacting with a \emph{many-electron} current,
thereby exploring quantum spin valve physics in the more-realistic regime beyond single-electron scattering.
%
%
The time-dependent density matrix renormalization group (td-DMRG) method \cite{schollwock_revmod, schollwock, chan_inpractice} 
is a well-established approach to the many-body transport problem \tddmrgtransport 
capable of simulating a many-electron scenario.
td-DMRG is a family of algorithms for representing, variationally minimizing, and time-evolving correlated many-body quantum states.
To study many-body transport, we apply the td-DMRG algorithm \cite{schollwock_revmod, schollwock, feiguin} to time evolve the initial many-body state under a quenched Hamiltonian using the time-dependent variational principle \cite{cirac_tdvp},
performing this calculation with the \textsc{block2} code \cite{block2}.
In these calculations, we find that an efficient many-electron quantum spin valve
is possible. 

\deled{and}\added{To further \inref{ciccarello}'s observation that
the quantum-spin valve effect}
hinges on enhancing the density of states at the Fermi energy,\deled{for the itinerant electrons.}
\added{our model Hamiltonian explicitly allows tunable band structure to probe how the underlying electronic properties of the nanowire impact the physics of the quantum spin valve.
Specifically, instead of a tight-binding model with fixed bandwidth as in \inref{ciccarello}, we use the Rice-Mele model to control the band structure, \textit{e.g.} the flatness of the band, \textit{via} the tight-binding parameters.
}
%
%
\added{The Rice-Mele model \cite{palyi, aachenthesis} is one-dimensional tight-binding model with staggered hopping
terms,
originally formulated to describe $\pi$-conjugated polymers \cite{ricemele, ssh, ssh_conjugated}.
The tunable staggered hopping terms enable Rice-Mele to capture the flat-band regime and tune other relevant electronic properties of the bands such as the  bandgap or density of states.
Drawing on $\pi$-conjugated \cite{chemically_conjugated_cnts, wernsdorfer_graphene} flat-band materials \cite{flatband_rev},
such as twisted bilayer graphene (TBG) \cite{macdonald, macdonald_rev} and its nanoribbon form \cite{tbg_nanoribbon}, doped graphene \cite{Cs_doped_graphene}, rhombohedral multilayer graphene \cite{rhombohedral_stacked_graphene, rhombohedral_acsnano},
and single wall carbon nanotubes (SWCNTs) \cite{roche_revmod, dresselhaus_swcnt, hydrogenated_cnts}
as exemplary realizations (Fig.~\ref{fig:setup_emul}) of $\pi$-conjugated nanowires with enhanced density of states, our results show that the quantum spin valve efficiency is greatly increased in the flat-band limit.
}

\section{Results and discussion}

\subsection{Model System}
\begin{figure}[t]
    \centering
    \includegraphics[width =\linewidth]{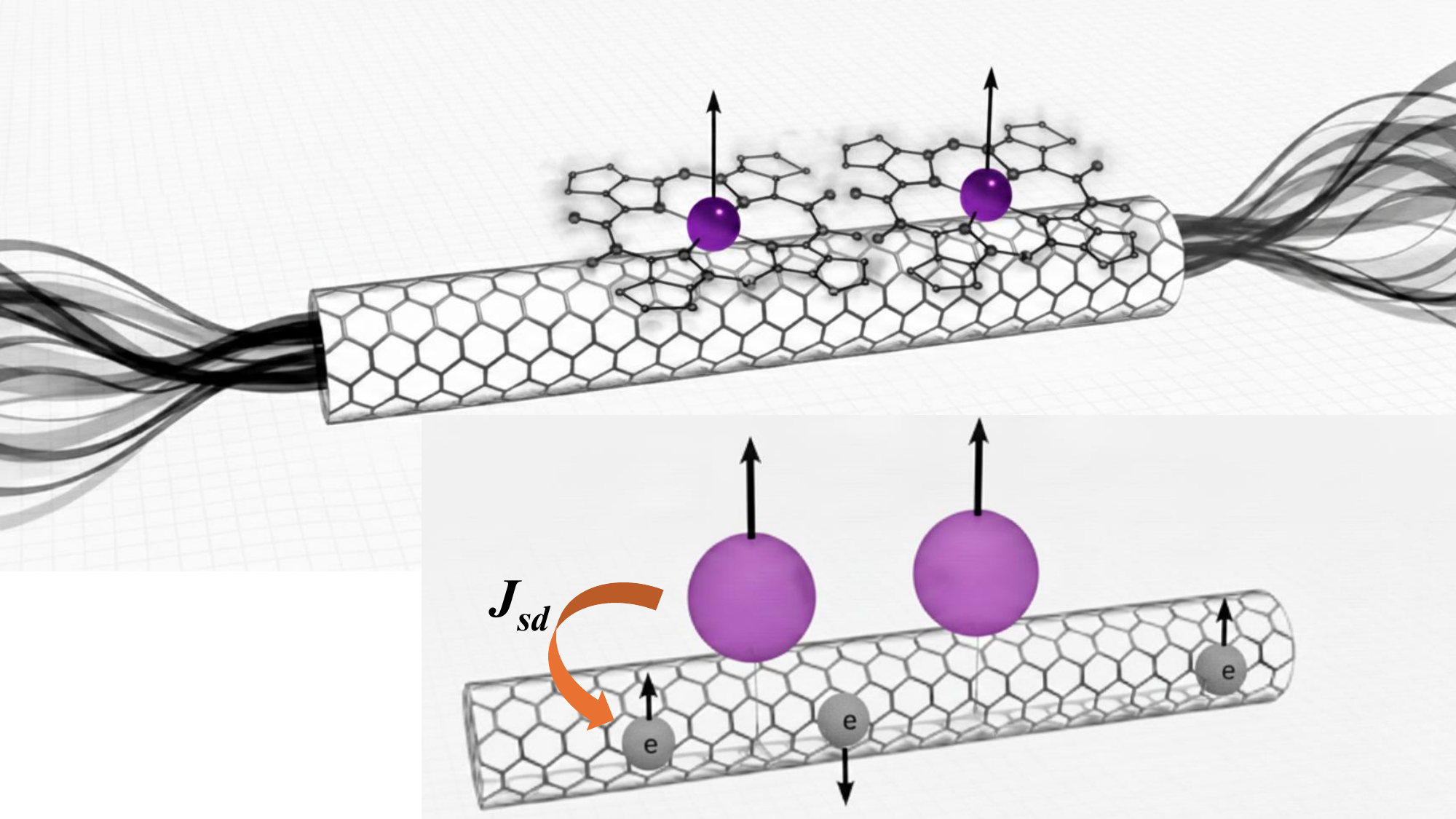}
    \caption[Supramolecular device with $sd$ exchange coupling itinerant electron spins and MSQs]{
    Top: \deled{In a}\added{Exemplary schematic of a} supramolecular device \added{where} MSQs (purple) retain their magnetic properties as they \deled{electronically couple to}\added{functionalize} a \deled{current-carrying}\added{$\pi$-conjugated} nanowire.
    Bottom: \deled{I}\added{Our model describes i}tinerant electrons in the nanowire \deled{see}\added{interacting with} the spin degree of freedom of the MSQs only, \deled{with}\added{\textit{via}} an exchange interaction \deled{$J_{sd}$}\added{of strength$J_{sd} \approx 1 $ meV} \deled{between itinerant electron spin and MSQ spin}. Image created with Gemini.
    }
    \label{fig:setup_emul}
\end{figure}

\begin{figure}[t]
    \centering
    \includegraphics[width =1.0\linewidth]{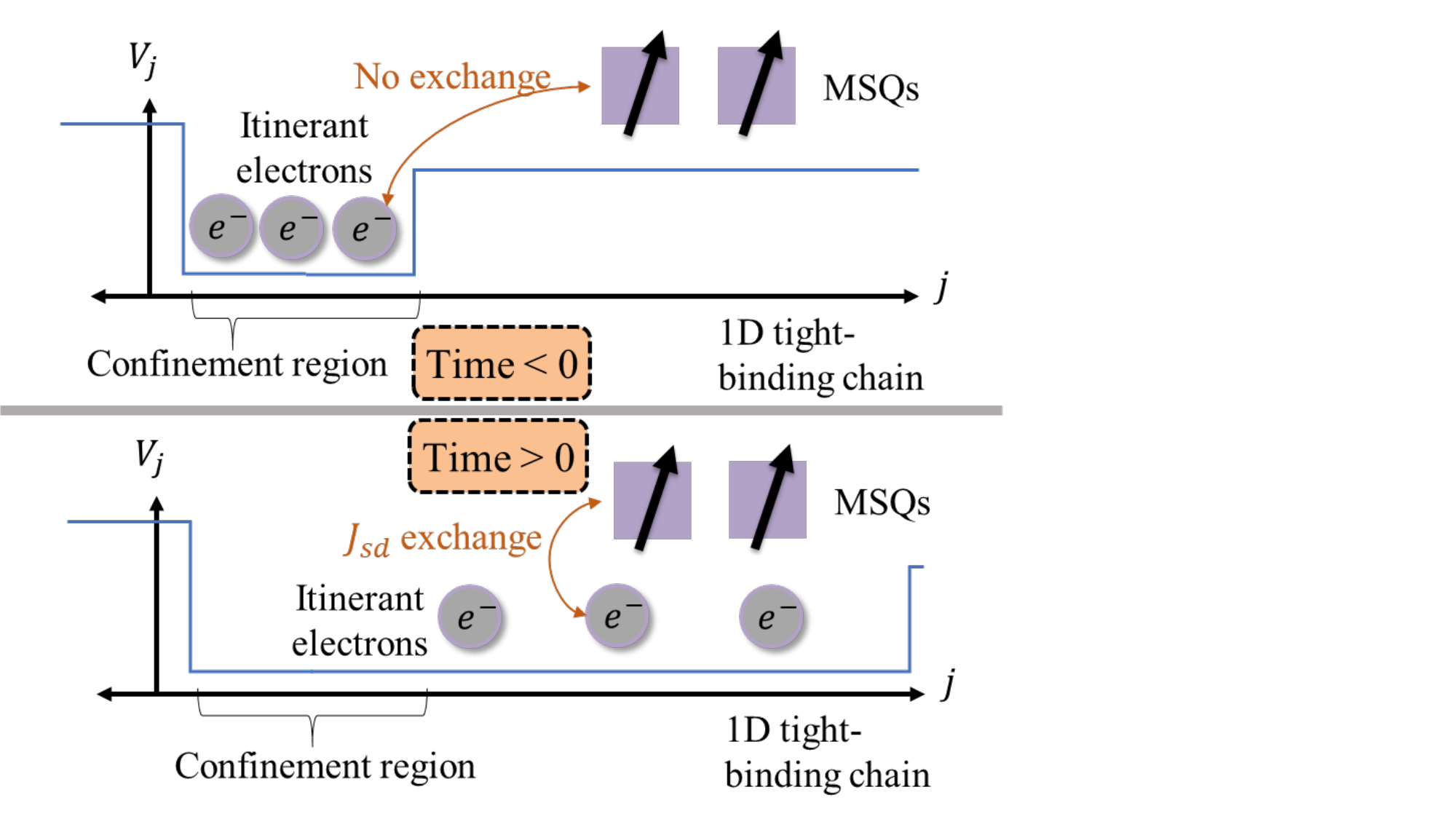}
    \caption[Model system]{Model system: itinerant electrons live in a finite 1D nanowire. After a quantum quench at time zero, they spread out and interact with MSQs (purple) atop the nanowire. When the itinerant electrons are close to the MSQs, virtual hopping onto the MSQs yields $sd$ exchange $J_{sd}$ but no Coulomb interactions. 
    }
    \label{fig:setup_model}
\end{figure}

%
\deled{The Hamiltonian for our model system is}\added{The model Hamiltonian for our system}
(Fig.~\ref{fig:setup_model}) \added{should couple the MSQs to itinerant electrons in the semiconducting nanowire and be flexible enough to capture different band structure regimes in the nanowire. We write down a model Hamiltonian with $sd$ exchange
and tunable band structure arising from Rice-Mele tight-binding terms:}

\begin{flalign}
    \hat{H}(t) =  \, v\sum_{j=0}^{N_\text{total}-1} &\sum_\sigma (c_{jA\sigma}^\dagger c_{jB\sigma} + c_{jB\sigma}^\dagger c_{jA\sigma}) \notag \\
    +w\sum_{j=0}^{N_\text{total}-2} &\sum_\sigma (c_{jB\sigma}^\dagger c_{j+1,A\sigma} + c_{j+1,A\sigma}^\dagger c_{jB\sigma}) \notag \\
    -\frac{J_{sd}}{\hbar} \sum_{j=N_L}^{N_L+N_{SR}-1} &\sum_{\sigma \sigma' \nu} \Big( S^\nu_{d=2j-2N_L} c_{jA\sigma}^\dagger \sigma^\nu_{\sigma\sigma'} c_{jA\sigma'} \notag \\
    &+ S^\nu_{d=2j+1-2N_L} c_{jB\sigma}^\dagger
    \sigma^\nu_{\sigma\sigma'}c_{jB\sigma'}  \Big) \notag \\
    + \theta(-t)\hat{H}_0\,,&
    \label{eq:H}
\end{flalign}
where
$0 \leq j \leq N_\text{total}-1$ discretizes the nanowire into \added{a one-dimensional chain of} tight-binding unit cells of width $a$,
and $\mu=A,B$ indexes the site degree of freedom within the cell.
Here $c_{j\mu\sigma}^{(\dagger)}$ destroys (creates) an electron with spin $\sigma \in [\uparrow, \downarrow]$ on site $\mu$ of unit cell $j$,
$\sigma^\nu$ are the Pauli matrices for $\nu=x,y,z$,
and there are $N_e$ total electrons.
The hopping from $A$ to $B$ sites in the same unit cell is $v$, while the hopping from $B$ to $A$ sites in the next unit cell is $w$.
The unit cells are partitioned into the left lead $0 \leq j \leq N_L-1$, the scattering region $N_L \leq j \leq N_L + N_{SR} - 1$, and the right lead $N_L+N_{SR} \leq j \leq N_\text{total}-1$.
For each site in the scattering region exactly one MSQ couples to the electronic spin occupying the site \textit{via} an exchange interaction $J_{sd}$.
\added{$J_{sd}$ describes itinerant electrons traversing the nanowire such that their occupation of the MSQs is suppressed, and Coulomb interactions therein are treated as virtual processes.}
The \added{MSQs are} spin-1/2\deled{ MSQs have}\added{s with} no spatial degrees of freedom, \deled{but are}\added{but are} described by their generators of rotation in spin space $S_d^\nu = \tfrac{\hbar}{2}\sigma_d^\nu$ for the $d^{th}$ MSQ. For the MSQ basis states we choose the eigenstates satisfying $S_d^z|\sigma_d\rangle = \pm \tfrac{\hbar}{2}|\sigma_d\rangle$ where $+(-)$ is for $\sigma = \uparrow(\downarrow)$.


 \deled{The staggered hopping values in Eq.~(\ref{eq:H}) represent a specific case of the Rice-Mele model \cite{ricemele, palyi, aachenthesis}.}\added{
The staggered Rice-Mele hopping terms $v$ and $w$ model the band structure with customizeable bandgap and bandwidth set by $w/v$ as discussed in the Supplementary Information. 
Setting
}
$v=-1$ \deled{fixes the}\added{ensures the overall} energy scale \deled{determined by}\added{agrees with} the material-specific bandwidth,
while \added{we vary} $w$ \deled{is variable}\added{to tailor the band structure}\deled{We survey $w$ values to model the Rice-Mele band structure (see the Supplementary Information ). Thus for our purposes (\ref{eq:H}) is a minimal tight-binding model of a system with two energy bands near the Fermi energy that we can deploy to customize the electronic properties, \textit{e.g.} the bandwidth, bandgap, or density of states at the Fermi energy, in our simulation to match the properties of real materials.}.
\added{
While the Rice-Mele model is the tool we have chosen to modify the band structure of the semiconductor nanowire and probe its effect on the quantum spin valve, any experimental realization of a quantum spin valve need not be described by Rice-Mele.
}

\deled{Finally}\added{In the last term in Eq.~(\ref{eq:H})}, $\theta(t)$ is the Heaviside function and $\hat{H}_0$ prepares the $t<0$ system state, confining the electrons and entangling the MSQs according to
\begin{flalign}
    &\hat{H}_0 = -V_\text{conf}\sum_{j=0}^{N_L-1} \sum_\mu \sum_\sigma c_{j\mu\sigma}^\dagger c_{j\mu\sigma} \label{eq:H0} \\
    &- \sum_{d=0}^{N_{SR}-1} \left( \frac{J_\text{ent}-iJ_\text{ent}'}{2\hbar^2}S_{2d}^+ S_{2d+1}^- 
    +\frac{J_\text{ent}+iJ_\text{ent}'}{2\hbar^2}S_{2d}^- S_{2d+1}^+ \right)
    \notag
\end{flalign}
where $S_d^\pm = S_d^x \pm i S_d^y$ and the $J_\text{ent}$, $J_\text{ent}'$ parameters determine $|\psi_\text{qubit}\rangle$ according to Eq.~(\ref{eq:define_phient}).
The term $\theta(-t)\hat{H}_0 $ equals zero after time zero, amounting to a \emph{quantum quench}
(for a review see \inref{quantum_quench})
of the system
under which the $t<0$ state of the system [the ground state of $\hat{H}(t<0)$]
is not an eigenstate of $\hat{H}(t>0)$ so the dynamics are nontrivial.
As sketched in Fig.~\ref{fig:setup_model},
removing the $V_\text{conf}$ term causes the itinerant electrons' wavefunctions spread out, simulating the injection of a current pulse into the nanowire \cite{feiguin}. 
Until it encounters the MSQs, the profile of this current pulse is determined by the single-particle eigenstates of $\hat{H}(t<0)$ occupied by the electrons, as discussed in 
the Supplementary Information
.
Using the Rice-Mele model for the electronic Hamiltonian allows us to explore different regimes (\textit{e.g.} a flat band structure) for these eigenstates that we would otherwise be unable to investigate.

\subsection{Results}
\label{sec:results}

\begin{figure*}[t]
    \centering
    \includegraphics[width =1.2\linewidth]{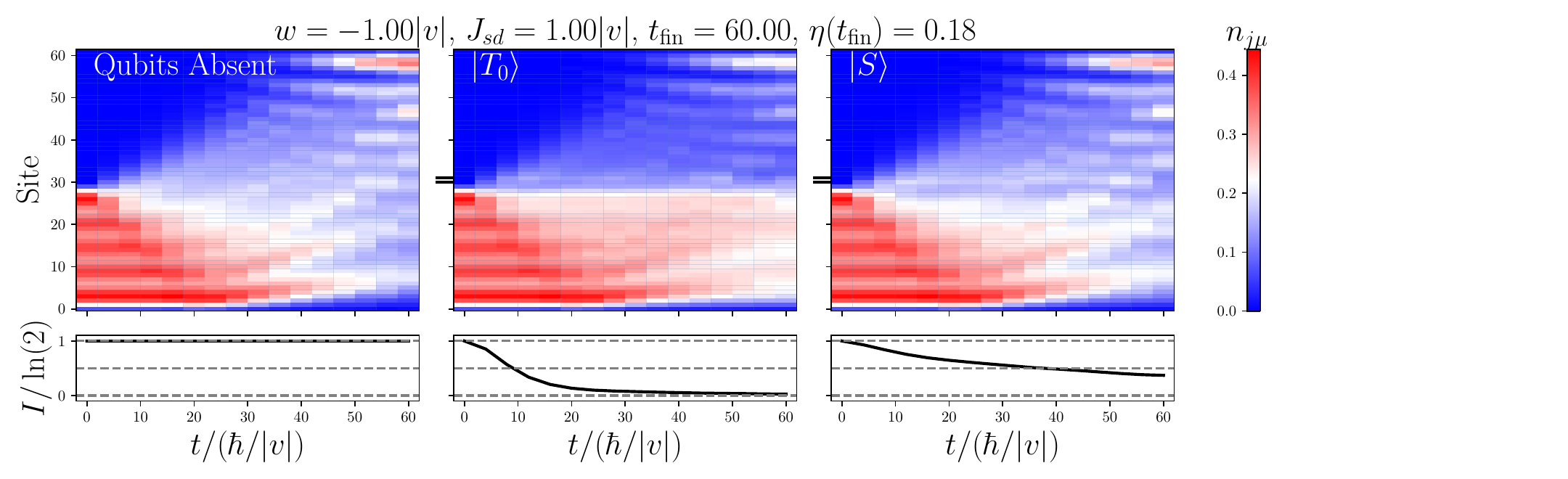}
    \caption{Panels show simulations of a system without MSQs, with $|T_0\rangle$ MSQs, and with $|S\rangle$ MSQs. 
    Upper plot of each panel:
    color shows the electronic occupation $n_{j \mu}$ of a given site at a given time. The sites $|j\mu\rangle$ discretizing the 1D chain are along the vertical axis while time $t$, measured in units $\hbar/|v|$, is along the horizontal axis. 
    Lower plot of each panel: perturbation of the MSQ entanglement, quantified by $I$ [Eq.~(\ref{eq:MI})].
    Hamiltonian parameters are $w=-1.00$, $J_{sd}=1.00$, $N_e = 10$, geometric parameters are $N_L=N_R=15$, $N_\text{total} = 31$, and td-DMRG parameters are $\chi_b = 250$, $dt=0.1$.} \label{fig:w-10}
\includegraphics[width =1.2\linewidth]{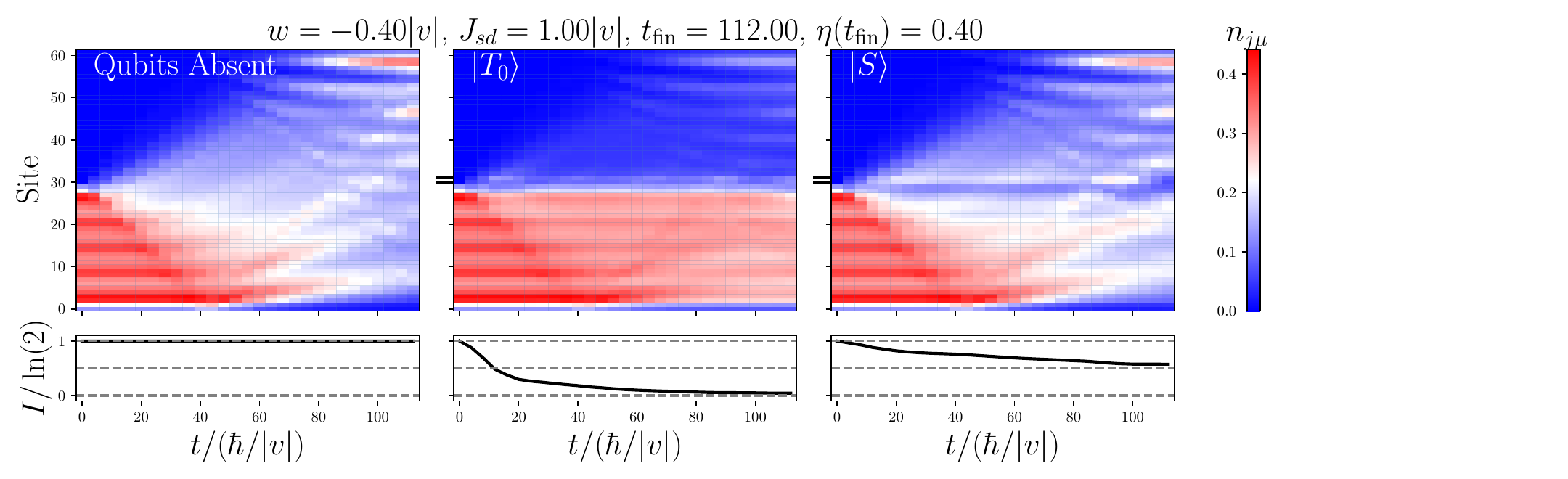}
    \caption{
    Panels show simulations of a system without MSQs, with $|T_0\rangle$ MSQs, and with $|S\rangle$ MSQs. 
    Upper plot of each panel:
    color shows the electronic occupation $n_{j \mu}$ of a given site at a given time. The sites $|j\mu\rangle$ discretizing the 1D chain are along the vertical axis while time $t$, measured in units $\hbar/|v|$, is along the horizontal axis. 
    Lower plot of each panel: perturbation of the MSQ entanglement, quantified by $I$ [Eq.~(\ref{eq:MI})].
    Notice the jump in quantum spin valve efficiency $\eta$ obtained by detuning $w$ from $w=-1.00$ to $w=-0.40$.
    Parameters besides $w$ are the same as Fig.~\ref{fig:w-10}.} 
    \label{fig:w-04}
\end{figure*}


Our finite system contains a left lead of $N_L=15$ unit cells, a scattering region of $N_{SR}=1$ unit cells containing two MSQs, and a right lead of $N_R = 15$ unit cells, with $N_e=10$ electrons confined to the left lead at $t<0$ by $\hat{H}_0$.
As stated in Model System the unit cell $j$ has site $\mu=A$ on the left and $\mu=B$ on the right, so that the 62 sites $|j\mu\rangle$ spanning the system fall into a one-dimensional chain.
The choice of 10 electrons, although far smaller than the current in many nanojunctions, 
has been experimentally achieved in STM experiments with picoampere currents applied for nanosecond pulses \cite{otte}.
Our results track the dynamics of this system under Eq.~(\ref{eq:H})
after the quantum quench removes
$\hat{H}_0$,
paying special attention to
the occupation $ n_{j\mu} = \sum_\sigma \langle c_{j\mu\sigma}^\dagger c_{j\mu\sigma} \rangle$ of each site $|j\mu \rangle$ along the one-dimensional chain
and mutual information $I$ between MSQs [Eq.~(\ref{eq:MI})] as a function of time $t$, which is measured in units of $\hbar/|v|$.

The quantum quench of the system initiates transport wherein the site occupation spreads rightward out of the left lead. This can be immediately seen
in our heatmap results displaying
the site occupation $n_{j \mu}$
as a function of time (Fig.~\ref{fig:w-10}).
We begin with the heatmap simulated when the MSQs are absent, making several remarks about the transport properties independent of the effects of $|\psi_\text{qubit}\rangle$.
First, after a certain time the electrons are observed to reflect off the right boundary $j=N_\text{total}-1$ of the finite system we are simulating.
The onset time for this reflection $t_\text{fin}$ demarcates when our finite-size simulations honestly depict the dynamics of an open system ($t<t_\text{fin}$) \cite{feiguin} from when our simulations no longer capture such dynamics ($t> t_\text{fin}$).
In Fig.~\ref{fig:w-10} finite-size effects can be seen at $t_\text{fin}=60$.
We present and discuss our definition of $t_\text{fin}$ in Eq.~(\ref{eq:tfin}).
We emphasize here that the timescale $t_\text{fin}$ is an important property of each transport simulation.

Second, to rigorously quantify the rightward transport for a given MSQ state, we measure
\begin{flalign}
    n_{R}(t, \phi_\text{ent}) =& \sum_{j=N_L+N_{SR}}^{N_\text{total}-1} \sum_\mu \sum_\sigma 
    \left \langle c_{j \mu \sigma}^\dagger c_{j \mu \sigma} \right \rangle \,,
    \label{eq:nR_alt}
\end{flalign}
the electron accumulation in the right lead at time $t$
when the phase of $|\psi_\text{qubit}\rangle$ [Eq.~(\ref{eq:cicc_state_alt})] is $\phi_\text{ent}$.
Our simulations are not accurate past $t=t_\text{fin}$
so we always report $n_R(t_\text{fin})$.
We note that experimentally,
$n_R(t_\text{fin})$ represents the integrated current through the junction over the window $[0,t_\text{fin}]$
and that an experimentalist could modulate the length of time the current is applied for to maximize the measured $n_R(t)$ signal.

To probe quantum spin valve physics, we compare heatmap results for transport
past initially entangled MSQ states in the family $|\psi_\text{qubit}\rangle$ defined by $\phi_\text{ent}$.
%
%
One descriptor of the MSQs throughout the transport is their mutual information $I$, which starts at $I=\ln(2)$ indicating maximal entanglement and decays with electron-MSQ interactions.
In Fig.~\ref{fig:w-10} we see that both $|T_0\rangle$ and $|S\rangle$ lose the majority of their entanglement
(see $I$ in the lower plot of each panel).
%
%
%
We also observe that the electrons pass the $|S\rangle$ MSQs and accumulate in the right lead, indicating an `open' quantum spin valve with good conductance.
For the single-electron transport in \inref{ciccarello} $|T_0\rangle$ becomes a corresponding `closed' quantum spin valve where the electron reflects off the MSQs rather than accumulating in the right lead.
We aim to replicate this disparity between $|S\rangle$ and $|T_0\rangle$ for many electrons,
however, 
the results for the trivial Rice-Mele band $w=v=-1$ (Fig.~\ref{fig:w-10}) the $n_{j\mu}$ heatmaps appear similar for both states. 
%
To rigorously define the difference in transport across MSQ states,
we introduce the quantum spin valve efficiency
\begin{flalign}
    \eta(t_\text{fin}) = \frac{
    n_R(t_\text{fin}, \phi_\text{ent}=\pi) - n_R(t_\text{fin}, \phi_\text{ent}=0)
    }
    {
    n_R(t_\text{fin}, \phi_\text{ent}=\pi) + n_R(t_\text{fin}, \phi_\text{ent}=0)
    }
    \,. \label{eq:efficiency_alt}
\end{flalign}
The efficiency quantifies how the conductance of the open state $|S\rangle$ differs from the closed state $|T_0\rangle$
analogously to the magnetoresistance of a classical spin valve \cite{baumgartner}.
For Fig.~\ref{fig:w-10} we calculate a relatively poor $\eta=0.18$ so this does not yet replicate an efficent many-electron quantum spin valve.

\begin{figure}
    \centering
    \includegraphics[width =1\linewidth]{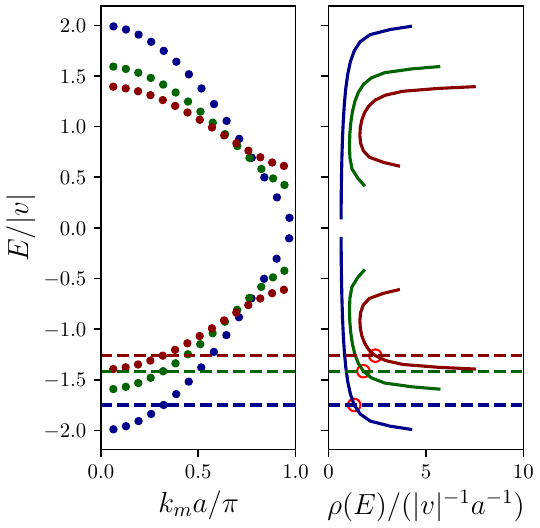}
    \caption[Discrete single-particle energy eigenstates of the finite nanowire]{Discrete $t>0$ single-particle energy eigenstates $E$ and associated wavenumbers $k_m$ and density of states $\rho(E)$. Spectrum computed for $w=-1$ (blue), $w=-0.60$ (green) and $w=-0.40$ (red) Rice-Mele bands.
    Dashed lines show  Fermi energy $E_F$ and red circles show $\rho(E_F)$. 
    }
    \label{fig:init_w}
\end{figure}

Having established that the results in Fig.~\ref{fig:w-10} fail to realize a high-efficiency quantum spin valve, we look at how varying the electronic properties can strengthen the efficiency.
At the level of our simulation, varying electronic properties means adjusting the Hamiltonian parameters, particularly $w$, in order to tailor the $t<0$ many-body state of the system,
$| \Psi(t<0)\rangle $ given in Eq.~(\ref{eq:t0manybody})].
For the electronic degrees of freedom, $| \Psi(t<0)\rangle $ comprises the single-particle eigenstates shown in
Fig.~\ref{fig:init_w} occupied by the itinerant electrons.
These single-particle eigenstates physically determine the energy and momentum of the electrons during transport.
In subsequent results, we change the Hamiltonian parameters, 
calculate $|\Psi(t<0) \rangle$ and the physically observable electronic properties such as $\rho(E_F)$, and finally determine the changes in the quantum spin valve simulations, particularly $\eta$.
Note that properties such as $\rho(E_F)$, arising from the Hamiltonian, are realistic as long as the Hamiltonian parameters are realistic
\footnote{
Given that we apply Rice-Mele, a two-band model, to emulate the band structure of a real material,
we ensure Hamiltonian parameters are reasonable by comparing features of our bands to real band structure.
For example, setting $|v|=$ 5 meV ensures that the bandwidth in our model when $w=0.4v$ agrees with the bandwidth reported for magic-angle twisted bilayer graphene \cite{vishwanath}.
}.

\subsection{Density of States Enhancement}

To strengthen the quantum spin-valve efficiency, we try customizing $\rho(E)$.
For a single band model (blue curve in Fig.~\ref{fig:init_w}) $\rho(E)$ only changes as the energy approaches the band edge, so density of states enhancement would have to proceed through ensuring itinerant electron energies are near the band edge, a challenging task of experimental control.
A simpler approach is flattening the bandwidth to enhance the density of states
across all energy states, which we can do within the Rice-Mele model by detuning $w$ in the regime $w<0$, $|w| < |v|$.
This should strengthen the quantum spin-valve efficiency,
has the additional advantage of
not relying on the precise number of electrons or their distribution in energy space.

Now taking $w=-0.40$ to flatten the bandwidth to $E_\text{band}=0.40$, 
in Fig.~\ref{fig:init_w} we observe that $\rho(E)$ is uniformly enhanced compared to values closer to $w=v$.
Looking at the electron accumulation for $|S\rangle$ vs. $|T_0\rangle$ in Fig.~(\ref{fig:w-04}),
thanks to the $\rho(E)$ enhancement the conductance dependence on the MSQ state can be seen by eye and the quantum spin valve efficiency has been increased to $\eta = 0.40$.
As with the $w=v$ simulation, $I$ is perturbed from its initially maximally entangled value $I=\ln(2)$ during the transport, but now we see that $|S\rangle$ retains more entanglement meaning it is less coupled to the itinerant electrons.

\subsection{Quantum Spin Valve}

\begin{figure}[t]
    \centering
    \includegraphics[width =1\linewidth]{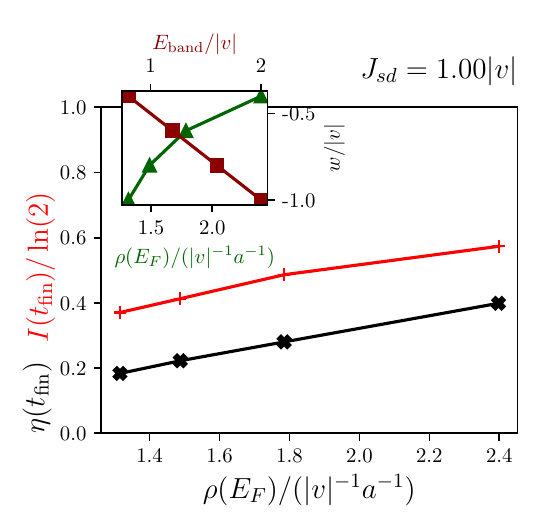}
    \caption[Quantum spin valve efficiency versus density of states]{Quantum spin valve efficiency $\eta$ [Eq.~(\ref{eq:efficiency_alt})] at the end of our simulations as a function of $\rho(E_F)$. 
    Entanglement retained by the $|S\rangle$ MSQs at the end of our simulations, quantified by the mutual information $I$ [Eq.~(\ref{eq:MI})].
    Both $\eta$ and $I$ correspond to a better quantum spin valve.
    Inset:
    the Rice-Mele hopping $w$ dictates one-to-one the bandwidth $E_\text{band}$ and the density of states at the Fermi energy $\rho(E_F)$.
    }
    \label{fig:efficiency}
\end{figure}

\begin{figure}[t]
    \centering
    \includegraphics[width =1\linewidth]{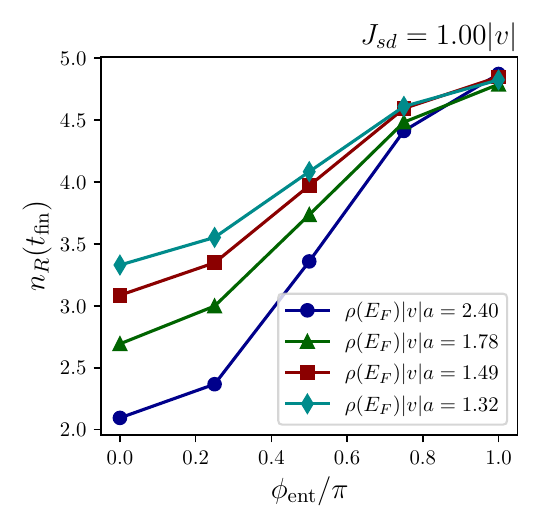}
    \caption[Sweeping over the entanglement phase to open the quantum spin valve]{Electron accumulation $n_R$ as we `open' the spin valve by sweeping the phase $\phi_\text{ent}$ of the entangled state (\ref{eq:cicc_state_alt}) from $\phi_\text{ent}=0$ to $\phi_\text{ent}=\pi$.
    Notice how $\rho(E_F)$ dictates the disparity between the closed and open valve.}
    \label{fig:vsphi}
\end{figure}

The increase in quantum spin valve efficiency from $\eta = 0.18$ when $w=-1.00$ in Fig.~\ref{fig:w-10} to $\eta=0.40$ when $w=-0.40$ in Fig.~\ref{fig:w-04} 
serves as a proof of concept for how using the Rice-Mele hopping to engineer electronic properties impacts the quantum spin valve efficiency.
In Fig~\ref{fig:efficiency} we show the full trend where enhancing the density of states at the Fermi energy $\rho(E_F)$ 
monotonically increases the efficiency of the quantum spin valve.
We can also look at the mutual information:
for an efficient quantum spin valve, $|S\rangle$ decouples more from the itinerant electrons and so should retain more entanglement between MSQs over the simulation. Within td-DMRG we can quantify the MSQ-MSQ entanglement with the mutual information $I$ and we see that $I$ also increases with $\rho(E_F)$.
The inset illustrates how
the Rice-Mele hopping $w$ directly determines the bandwidth and $\rho(E_F)$. 

Beyond the efficiency, we can also explore how $\rho(E_F)$ impacts the detailed dependence of the conductance on all the entangled states in the family (\ref{eq:cicc_state_alt}).
In Fig.~\ref{fig:vsphi}, we sweep over the values of $\phi_\text{ent}$ defining the maximally entangled states between $|T_0\rangle$ ($\phi_\text{ent}=0$) and $|S\rangle$ ($\phi_\text{ent}=\pi$).
Remarkably, our quantum spin valve is not just a bimodal device but \emph{continuously tunes} its conductance from lower to higher as we sweep $\phi_\text{ent}$.
Further, we see increasing $\rho(E_F)$ increases the gradient across this sweep just as it increases $\eta$.

\subsection{Qubit Read-Out}

\begin{figure}[t]
    \centering
    \includegraphics[width =1\linewidth]{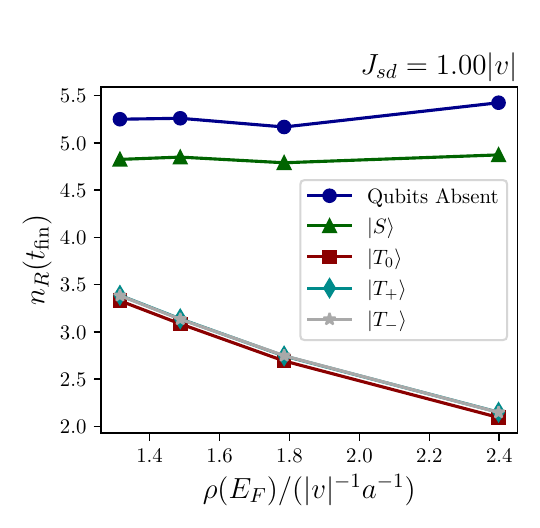}
    \caption[Singlet vs triplet family of states]{Electron accumulation $n_R$  for $|S\rangle$ vs. all triplet states as a function of density of states at the Fermi energy $\rho(E_F)$.
    Enhancing $\rho(E_F)$ suppresses the conductance for all triplet states in tandem,
    making the $|S\rangle$ behavior stand out more clearly.}
    \label{fig:transport_metric}
\end{figure}

For read-out of $ST_0$ qubits the full Hilbert space includes the maximally entangled states $|S\rangle$, $|T_0\rangle$ and the unentangled states $|T_+\rangle = |\uparrow_1 \uparrow_2 \rangle $, $|T_-\rangle = |\downarrow_1 \downarrow_2 \rangle $ .
A conventional Pauli spin blockade experiment \cite{petta, loss_tarucha, nichol_prapp, ladd_PSB} shows an electrical signal as a qubit charge tunnels for the $|S\rangle$ state but no such signal for the $|T_\pm\rangle$ or $|T_0\rangle$ states, distinguishing $|S\rangle$ from the triplet family of states.
We now ask whether we can apply the quantum spin valve physics to make the same distinction for MSQs.
This would allow a measurement of the nanowire conductance,
which has already been applied to experimental read-out of spin states in molecules \cite{wernsdorfer_nuc, wernsdorfer_nuc2, wernsdorfer_msq},
to replace Pauli spin blockade in reading-out molecular $ST_0$ qubits.
This replacement is desirable since
MSQs lack the tunneling control characteristic of semiconductor spin qubits in gate-defined quantum dots \cite{nichol_prapp}, a necessary condition to achieve the Pauli spin blockade tunneling signal \cite{wernsdorfer_msq}
\footnote{
In fact, the latest high-fidelity Pauli spin blockade schemes involve charge-latching \cite{nichol_prapp} which requires even more electron tunneling control
}.

In Fig.~\ref{fig:transport_metric}, we illustrate the electron accumulation [Eq.~(\ref{eq:nR_alt})] results for $|S\rangle$ vs. the entire family of triplet states.
Remarkably, we see that not only does the density of states enhancement separate the measured accumulation of $|S\rangle$ from $|T_0\rangle$ as we have seen with the quantum spin valve, it distinguishes all the triplet states from $|S\rangle$.
A conductance measurement for the two MSQs deposited atop the nanowire thus stands as a suitable analogue of Pauli spin blockade that allows read-out of $ST_0$ qubits encoded in molecules.
The electrical signal of high conductance (high electron accumulation) distinguishes $|S\rangle$ from  $|T_\pm\rangle$ and $|T_0\rangle$.
Of course, in any real experiment one wants to maximize the strength of this signal to increase the fidelity of assigning a measurement to a singlet or triplet state.
Our results show that 
\emph{the density of states enhancement serves as the physical knob for optimizing the read-out fidelity}
through increasing the quantum spin valve efficiency.
Although in this paper we used the Rice-Mele hopping to tune this knob, an experimenter is not limited to systems with staggered hopping but must simply be able to enhance $\rho(E_F)$.

\subsection{Discussion}

%
%
\added{
The supramolecular devices \supramolecular we model in this work consist of MSQs that retain their magnetic properties when noncovalently functionalizing \noncovalent $\pi$-conjugated nanowires.
Our model assumes (i) a $\pi$-mediated $sd$ exchange between the itinerant electrons of the $\pi$-conjugated nanowire and the MSQs and (ii) flat-band electronic dispersion in the nanowire.
%
%
%
We now discuss whether the components of this model are manifest in the experimental literature.
}

\added{
For a chemical realization of the MSQs, we propose metal phthalocyanines (MPcs), which offer a range of qubit properties through choice of the metal, such as an electronic spin-1/2 with microsecond coherence times for VOPc \cite{sessoli_vopc} or a nuclear spin-3/2 with coherence times on the millisecond scale for TbPc$_2$ \cite{wernsdorfer_algo}.
The facile attachment of MPcs to $\pi$-conjugated surfaces such as graphene and carbon nanotubes via aromatic ligands \noncovalent
creates supramolecular devices (Fig.~\ref{fig:setup_emul})
of itinerant electrons coupled to MSQs without fundamentally altering either \cite{chemical_reviews}.
An alternative, but less well-studied, device might feature polynuclear single molecule magnets such as Mn$_{12}$ functionalized to the nanowire by electronegative ligands \cite{liu_Mn12, christou_hebard}.
}

\added{
Chemical realization of the $sd$ exchange intrinsically depends on the MSQ and its surface functionalization,
because an electronic pathway between the itinerant electrons and MSQs is a prerequisite for $sd$ exchange \cite{wernsdorfer2}.
For the MPc example, the spin of the Pc macrocycle couples the qubit to its environment \cite{wernsdorfer_spin_hamiltonian, smms_fm_substrates, Pc_pi_orbitals, wernsdorfer_gmr}.
In the case of $\pi$-conjugated nanowires, this pathway is $\pi$-mediated \cite{wernsdorfer2, relay_like_dont_cite_Jsd}.
For the exemplary system in Fig. ~\ref{fig:setup_emul} this pathway \cite{wernsdorfer2} may stretch from MSQ ligands, to aromatic anchoring groups such as hexyl \cite{acsnano_2010, wernsdorfer2, wernsdorfer_graphene} pyrenyl \cite{acsnano_2010,wernsdorfer2, wernsdorfer_graphene, jacs_2009, jacs_2010} or its derivatives \cite{jphyschemC_2007},
to the $\pi$-conjugated nanowire.
An appropriate value for the coupling constant in such systems is $J_{sd}\approx 1$ meV \cite{wernsdorfer2}, many orders of magnitude smaller than the hopping amplitude in graphene \cite{roche_revmod}.
In contrast, in our work we have set $|v|$ and $J_{sd}$ to the same order of magnitude, which is justified in the case that the nanowire is a flat-band material, \textit{e.g.} TBG for which a bandwidth of 4 meV \cite{vishwanath} yields $|v|=5$ meV in the $w=-0.4v$ case.
}

\added{
Inspired by the electronic properties of $\pi$-conjugated materials in supramolecular devices, to describe itinerant electrons in the nanowire we wrote down a one-dimensional tight-binding model of the Rice-Mele form that can access different regimes for the band flatness, bandgap, and density of states.
However, we stress that experimental realizations of flat-band electrons are in fact more complicated than the one-dimensional tight-binding system in our simulations.
For a chemical realization of the nanowire, the $\pi$-conjugated materials mentioned as ideal for MPc functionalization yield plentiful examples of flat-band physics, 
including TBG \cite{macdonald_rev, macdonald} with bandwidth of a few meV \cite{vishwanath},
zigzag SWCNTs \cite{roche_revmod} under a magnetic field \cite{dresselhaus_swcnt} or under hydrogenation \cite{hydrogenated_cnts} with bandwidth of hundreds of meV, and doped graphene with a bandwidth about ten meV \cite{Cs_doped_graphene}.
Recalling that our simulations established large density of states near the Fermi energy as the desirable property,
we can tune $\rho(E_F)$  in TBGs \textit{via} the twist angle \cite{macdonald_rev, macdonald} or in doped graphene by substituting the dopant \cite{Cs_doped_graphene}.
} 
\deled{
For a physical realization of the nanowire, we need a 1D material with coherent electron transport.
Single-wall carbon nanotubes (SWCNTs) are such materials \cite{tans} with the additional benefit of secure attachment to MSQs \cite{wernsdorfer2}.
Since reading-out the $ST_0$ state requires isolating the dependence of the conductance on $|\psi_\text{qubit}\rangle$,
the nanowire conductance should not have contributions from thermally activated carriers \cite{mceuen_thermal},
which would be so large as to wash out differences in the conductance arising from the $|S\rangle$ vs. $|T_0\rangle$.
The majority of zigzag SWCNTs are semiconducting
with band gaps on the order 500 meV, so there will be no thermal activation at room temperature.
The final nanowire criterion, evident from our simulation of an efficient many-electron quantum spin valve in Fig.~\ref{fig:efficiency},
is that we should be able to modulate the band structure and enhance the density of states at the Fermi energy,
as we were able to do at the level of simulation by detuning $w$ (inset of Fig.~\ref{fig:efficiency}).
}
\added{Likewise,} SWCNTs have widely tunable band structures that can be controlled by various levers during fabrication, or \textit{in situ} by doping or application of electric or magnetic fields
\cite{qian_band_engineering}.\deled{Considering only semiconducting zigzag SWCNTs unlocks extensive tunability, as the tube circumference $R_{circ}$ modifies the bandgap by a factor $1/R_{circ}$ \cite{roche_revmod}
or scales the density of states curve by the same factor \cite{mintmire_universal}.

To translate our results, engineering enhanced $\rho(E_F)$ is the key bottleneck.
For semiconducting zigzag SWCNTs,
a magnetic field perpendicular to the tube axis reduces the bandwidth of the bands immediately above and below the Fermi energy until they become flat, enhancing the density of states \cite{dresselhaus_swcnt}.
Note that the magnetic field strength required to achieve this increases
as the tube circumference decreases
\cite{roche_revmod}.
Alternatively, for so-called `metallic' zigzag SWCNTs characterized by a band gap an order of magnitude smaller than semiconducting zigzag SWCNTs \cite{lieber},
the density of states at the Fermi energy can be enhanced linearly with the inverse tube circumference \cite{mintmire_universal} or exponentially with the strength of a magnetic field applied perpendicular to the tube \cite{ando_seri, roche_revmod}.}
For semiconducting zigzag SWCNTs,
\added{altering the hydrogenation \cite{hydrogenated_cnts} or applying} a magnetic field perpendicular to the tube axis \added{\cite{dresselhaus_swcnt}} \deled{reduces the bandwidth of}\added{enhances $\rho(E_F)$ by flattening} the bands immediately above and below the Fermi energy.
Alternatively, for so-called `metallic' zigzag SWCNTs characterized by a band gap an order of magnitude smaller than semiconducting zigzag SWCNTs \cite{lieber},
the density of states at the Fermi energy can be enhanced linearly with the inverse tube circumference \cite{mintmire_universal} or exponentially with the strength of a magnetic field applied perpendicular to the tube \cite{ando_seri, roche_revmod}.

\subsection{Conclusion}


\added{To conclude,
our results show that an efficient quantum spin-valve effect occurs not only for single-electron scattering but extends to many-electron current pulses in real devices.
We demonstrated that the quantum spin valve identified in \inref{ciccarello} is recovered with many electrons as long as the density of states at the Fermi energy is enhanced (Fig. \ref{fig:efficiency}), which can be accomplished
by detuning $w$ (inset of Fig. \ref{fig:efficiency}).
We used the Rice-Mele model to elucidate this essential connection between electronic properties and quantum spin valve efficiency by capturing the salient features of the first conduction and valence bands, \textit{e.g.} bandwidth. However, the Rice-Mele model or any minimal two-band model cannot capture the full electronic structure of $\pi$-conjugated surfaces such as TBG, SWCNTs or graphene. 
Besides the obvious elision of bands further away from the Fermi energy, 
Rice-Mele fails to capture curvature effects \cite{roche_revmod} and the chirality of the Dirac cones \cite{vishwanath}.
An atomistic tight-binding model of the $\pi$-orbitals of the surface would provide a more sophisticated description of the electronic structure \cite{roche_revmod, macdonald, vishwanath}.
%
%
Our survey of experimental work illustrates that candidate graphene-derived nanowires exist that can be coupled to MSQs, exhibit flat bands, and experimentally enhance $\rho(E_F)$, demonstrated in our results to correlate with an efficient quantum spin valve.
}




\section{The Time-Dependent Density Matrix Renormalization Group Method}
\label{sec:td-dmrg}


Since we aim to explore the functionality of MSQs in a current-carrying device, we have two perogatives for our methods. First, since we are describing qubits, our methods must be fully quantum mechanical. Second, since the qubits are coupled to the nanowire for read-out, our methods must be scalable enough to tackle this extended system. 
The combination of these poses a significant methodological challenge.
The power of td-DMRG is that we can explore larger systems while retaining a fully quantum description of the dynamics. Many works have already taken advantage of this to investigate transport in nanoscale devices and reveal quantum effects \tddmrgtransport. However, when it comes to entanglement, the story becomes more complicated.
With continued progression towards qubit read-out, built around discrete, controllable quantum systems, it becomes increasingly important to pick out entanglement between specific, spatially differentiated subsystems.
Whereas previous works only measure the entanglement that develops between all of the subsystems in aggregate,the approach we describe in Sec.~\ref{sec:sp_entanglement} incorporates from the outset the ability to resolve entanglement between any chosen degrees of freedom in the position domain.
By choosing an entanglement metric with this level of spatial resolution, we capture entanglement
\emph{between each individual molecular spin along the nanowire,}
and can therefore interrogate the function of these objects as qubits in a device.

\subsection{Time Dependence}

The starting point of time dependent DMRG is the ansatz for the many-body state, the so-called matrix product state \cite{chan}. We detail the construction and advantages of the matrix product state in 
the Supplementary Information.
For td-DMRG, we need an efficient way to variationally minimize the many-body state, calculate expectation values from it, and unitarily time evolve it.
The matrix product state unlocks this efficiency by way of its tensor structure,
using matrices $\textbf{A}^{(l)n_l}$ of dimension $\chi_b$ (refered to in the literature as the bond dimension) at each site $l$ in the $L$ site system to create an $\mathcal{O}(\chi_b^2 L)$ variational parameter approximation
of the $4^L$ parameter exact many-body state.
The tensor structure of the matrix product state allows efficient computation of state overlaps as well as one and two-body expectation values \cite{chan}.
For variational minimization, we employ the two-site sweep algorithm \cite{chan_inpractice} from \textsc{block2} \cite{block2}. This consists of sweeping over subsequent pairs of matrices $\textbf{A}^{(l)n_l} \textbf{A}^{(l+1)n_{l+1}}$ appearing in Eq.~(\ref{eq:mps}) and varying the matrix elements of that pair while keeping all others fixed
\cite{chan_inpractice, white_algo, tn_dmrg}.
Sweeping over $\textbf{A}^{(l)n_l} \textbf{A}^{(l+1)n_{l+1}}$ pairs remains more robust against local minima than simply varying the matrix elements of a single $\textbf{A}^{(l)n_l}$ each time
\cite{chan_inpractice}.

To evolve in time by a step $dt$, we operate on the matrix product state with $\exp(-i\hat{H}dt/\hbar)$.
This operator must be decomposed,\textit{e.g.} by Suzuki-Trotter \cite{feiguin, feiguin_trotter}.
Suzuki-Trotter decompositions perform poorly for Hamiltonians with long-range interactions \cite{feiguin_tst}
and sometimes fail to obey all conservation laws and reflect all the symmetries of the system.
To remedy these deficiencies, other DMRG time evolution methods have been developed.
One method is time step targeting,
which expands on Suzuki-Trotter to optimize the matrix product state basis for the current time step and next time step simultaneously for more accurate treatment of long range interactions
with controlled
\footnote{
Controlled in the sense that we suffer the computational cost of covering only one additional time step compared to Suzuki-Trotter.
}
loss of efficiency \cite{feiguin_tst}.
Another method is the time-dependent variational principle \cite{cirac_tdvp, shuai}, which conserves energy and all other constants of motion throughout.
Here, we use the time-dependent variational principle for time evolution, leading to computational complexity that scales as $\mathcal{O}(\chi_b^3)$ \cite{cirac_tdvp}
which happens to be the same scaling as the variational determination of the ground state matrix product state
\cite{chan_inpractice}.
In brief, the matrix product state is written in terms of time-dependent matrices $\textbf{A}^{(l)n_l}(t)$ and the time-dependent variational principle dictates that the time derivatives of these matrices satisfy symplectic equations conserving all constants of motion \cite{cirac_tdvp}. Then, integrating these equations tells the time evolution of the state.
Again, we emphasize that approximating the many-body state and time evolution with a matrix product state
expands the terrain we can cover with our computational resources, as exact time evolution of a quantum system of the size and for the length of time we are interested in would be prohibitively expensive.

\subsection{Spatially Resolved Entanglement in td-DMRG}
\label{sec:sp_entanglement}

The matrix product state approximation of the quantum state assumes that the full system can be represented by a \emph{pure} density matrix
\begin{flalign}
    \rho_\text{MPS} = |\Phi_\text{MPS}\rangle \langle \Phi_\text{MPS} | \,
\end{flalign}
where MPS abbreviates matrix product state.
$\rho_\text{MPS}$ is defined over all $L$ orbitals of the system and can be used
to compute reduced density matrices of the subsystems.
The reduced density matrix of orbital 1 is defined as
$ 
    \rho_1 = \text{Tr}_{2,3,4,...}[\rho_\text{MPS}] \,,
$ 
the reduced density matrix  of orbitals $1$ and $2$ combined is
$ 
    \rho_{12} = \text{Tr}_{3,4,...}[\rho_\text{MPS}] \,,
$ 
and so on. Due to interactions between the orbitals, the reduced density matrices are in general mixed even though $\rho_\text{MPS}$ is pure.

The von Neumann entropy of an arbitrary (pure or mixed) density matrix $\rho_\text{arb}$ is \cite{solyom, wootters, wootters_prl}
\begin{flalign}
    S(\rho_\text{arb}) = -\text{Tr}[\rho_\text{arb} \ln(\rho_\text{arb})] = -\sum_\alpha \lambda_\alpha \ln(\lambda_\alpha) \,,
    \label{eq:VNE}
\end{flalign}
where the second equality comes from extracting the eigenvalues $\lambda_{\alpha}$ from $\rho_\text{arb}$ \myeq{reiher}{2}. The von Neumann entropy of a density matrix quantifies the entanglement between (i) all the degrees of freedom included in that density matrix and (ii) all the other degrees of freedom of the system. It does not tell us anything about entanglement between the degrees of freedom included in the density matrix. The von Neumann entropy of a pure state is zero.

In this work, we study MSQs in the context of a current-carrying device, focusing on how the MSQ entangled states modulate the device conductance.
\emph{This requires a method powerful enough to simulate the qubits and the rest of the device interacting together, but retaining the resolution to distinguish between the qubits and the device.}  
Many previous applications \cite{feiguin, shuai, burrello, feiguin_wilsonchain}
of td-DMRG to the transport problem have studied the entanglement of the system as a whole, but lacked detailed exploration of the entanglement between specific subsystems. 
When considering MSQs in a device, these finer details are essential, as we want to track entanglement between MSQs and avoid entangling them with the rest of the device
\footnote{
Sometimes, \textit{e.g.} read-out, we do aim to entangle the MSQs with a specific subsystem. But entanglement with the rest of the device in general is always undesirable.
}.
In this section we illustrate how to quantify entanglement with this level of granularity within td-DMRG.

We now calculate the entanglement between exactly two molecular orbitals, \textit{e.g.} $l=1$ and $l=2$, of a system comprising $L>2$ molecular orbital subsystems. This is what we mean by `spatially resolved entanglement,' \textit{i.e.} isolating the entanglement \textit{between} 
a pair of degrees of freedom that differ in their spatial location,
while leaving out any entanglement they may possess with the rest of the system. To do this we determine the reduced density matrices of molecular orbitals 1 and 2 ($\rho_1$ and $\rho_2$),
then calculate the von Neumann entropy of these individually [$S(\rho_1)$ and $S(\rho_2)$] and jointly [$S(\rho_{12})$], and finally extract the mutual information 
\myeq{reiher}{3}
\begin{flalign}
    I(l,l') = \frac{1}{2}\left[S(\rho_{ll'}) - S(\rho_l) - S(\rho_{l'}) \right](1-\delta_{lm})\,.
    \label{eq:MI}
\end{flalign}
The mutual information is zero for unentangled subsystems and grows as the entanglement grows.

Specializing now to $\rho_\text{MPS} $, the $l^{th}$ molecular orbital is spanned by $\{ |0\rangle_l, |\uparrow\rangle_l, |\downarrow\rangle_l, |\uparrow\downarrow\rangle_l \}$ and its reduced density matrix is formally obtained as \cite{white_VNE}
\begin{flalign}
    \rho_l =& \text{Tr}_{1,...l-1,l+1,...L} \left[ |\Phi_\text{MPS} \rangle \langle \Phi_\text{MPS} | \right] \,.
    \label{eq:rho_1orb}
\end{flalign}
The problem we face is that there is no way to do the required trace operations for an matrix product state. Fortunately, all the matrix elements of $\rho_l$ can be formulated as expectation values of strings of
$c_{l\sigma}^{(\dagger)}$ operators with respect to $|\Phi_\text{MPS}\rangle$, which are easy to compute in td-DMRG since $|\Phi_\text{MPS}\rangle$ is always known.
This procedure yields the closed form expression \mycite{white_VNE}{Table 1}
\begin{flalign}
    &\rho_l = \text{Tr}_{1,...l-1,l+1,...L} \left[ |\Phi_\text{MPS}\rangle \langle \Phi_\text{MPS} | \right] = \label{eq:1ord_rdm}  \\
    &
    \begin{pmatrix}
         \langle \eta_{ll} \rangle & 0 & 0 & 0 \\
         0 & \langle n_{l\uparrow} (1-n_{l\downarrow})\rangle & 0 & 0 \\
        0 & 0 & \langle n_{l\downarrow} (1-n_{l\uparrow})\rangle & 0 \\
        0 & 0 & 0 & \langle n_{l\downarrow} n_{l\uparrow}\rangle \rangle \\
    \end{pmatrix}
    \begin{pmatrix}
        |0\rangle_l \\
        |\uparrow\rangle_l \\
        |\downarrow \rangle_l \\
        |\uparrow \downarrow \rangle_l
    \end{pmatrix}
    \notag
\end{flalign}
where $n_{l\sigma} \equiv c_{l\sigma}^\dagger c_{l\sigma}$ and $\eta_{ll'} = (1-n_{l\downarrow})(1-n_{l'\uparrow})$. 
In practice, we always apply this to calculate the von Neumann entropies of different MSQs along the nanowire. Since the MSQ subspace is $\{ |\uparrow\rangle_d, |\downarrow\rangle_d\}$ [Eq.~(\ref{eq:supersiting})], we only need to calculate the central $2\times 2$ block of Eq.~(\ref{eq:1ord_rdm}).

For two orbital reduced density matrices, we again sidestep tracing over $\rho_\text{MPS}$ in favor of a closed form expression that depends only on expectation values. The expression for the full $16 \times 16$ reduced density matrix is given in  given in Tables 3 and 4 of Boguslawski et al. \cite{reiher}, but as with Eq.~(\ref{eq:1ord_rdm}) we truncate from molecular orbitals to spin orbitals, yielding a $4\times 4$ matrix with nonzero elements 
\begin{flalign}
    \langle \uparrow_l \uparrow_{l'} |\rho_{l,l'} |\uparrow_l \uparrow_{l'} \rangle =& \langle (1-n_{l\downarrow}) n_{l\uparrow}(1-n_{l'\downarrow})n_{l'\uparrow}\rangle \notag \\
    \langle \uparrow_l \downarrow_{l'} |\rho_{l,l'} |\uparrow_l \downarrow_{l'} \rangle =& \langle(1-n_{l\downarrow})n_{l\uparrow}(1-n_{l'\uparrow})n_{l'\downarrow}\rangle \notag \\
    \langle \uparrow_l \downarrow_{l'} |\rho_{l,l'} |\downarrow_l \uparrow_{l'} \rangle =& \langle-c_{l\downarrow}c_{l\uparrow}^\dagger c_{l'\downarrow}^\dagger c_{l'\uparrow} \rangle \notag \\
    \langle \downarrow_l \uparrow_{l'} |\rho_{l,l'} |\uparrow_l \downarrow_{l'} \rangle =& \langle-c_{l\downarrow}^\dagger c_{l\uparrow} c_{l'\downarrow} c_{l'\uparrow}^\dagger \rangle \notag \\
    \langle \downarrow_l \uparrow_{l'} |\rho_{l,l'} |\downarrow_l \uparrow_{l'} \rangle =& \langle(1-n_{l\uparrow})n_{l\downarrow}(1-n_{l'\downarrow})n_{l'\uparrow}\rangle \notag \\
    \langle \downarrow_l \downarrow_{l'} |\rho_{l,l'} |\downarrow_l \downarrow_{l'} \rangle =& \langle(1-n_{l\uparrow}) n_{l\downarrow}(1-n_{l'\uparrow})n_{l'\downarrow}\rangle
    \label{eq:2orb_rdm}
\end{flalign}

We use Eqs.~(\ref{eq:1ord_rdm})-(\ref{eq:2orb_rdm}) to calculate the requisite reduced density matrices 
and diagonalize them to find the von Neumann entropies, finally arriving at the mutual information through Eq.~(\ref{eq:MI}). $I(l,l')$ quantifies the spatially resolved entanglement between the different MSQs of the system, varying from zero for unentangled MSQs to $\ln(2)$ for maximally entangled MSQs. 
We note that computing $\rho_{l,l'}$ in this way also enables computation of other useful metrics of qubit-qubit entanglement such as the negativity \cite{vidal_negativity}.

\begin{acknowledgement}

We appreciate the plentiful guidance H. Zhai and G. Chan provided so that we could fully utilize their \textsc{block2} td-DMRG code \cite{block2}. We also thank A. Feiguin for many helpful discussions about td-DMRG, its strengths and weaknesses, and its application in quantum quench and quantum transport problems.
This work was supported as part of the Center for Molecular Magnetic Quantum Materials, an Energy Frontier Research Center funded by the U.S. Department of Energy, Office of Science, Basic Energy Sciences under Award no. DE-SC0019330.
Computations were done using the utilities of the National Energy Research Scientific Computing Center and University of Florida Research Computing and some graphics were created using Gemini.

\end{acknowledgement}

\begin{suppinfo}

In the following supporting information, we give details on the Rice Mele model
and explain the derivation of the matrix product state ansatz used for the td-DMRG simulations.


\section{Rice-Mele Model}
\label{app:RiceMele}

\newcommand{\dummy}
{ 
\begin{figure*}[t]
    \centering
    \includegraphics[width =1\linewidth]{figures/dispersion_RiceMele.pdf}
    \caption{Detuning $w$ narrows the bandwidth of the energy $E$ vs. wavenumber $k$ bands and enhances the density of states $\rho$.}
    \label{fig:dispersion_RiceMele}
\end{figure*}
}

The Rice-Mele model is a minimal model of a system with two energy bands admitting a rich variety of band structures as seen in Fig.~\ref{fig:init_w}.
In general, the Rice-Mele model consists of staggered real hopping parameters $v$ (intracell) and $w$ (intercell), as well as staggered intracell potential $u$. 
Rather than fitting these to real-space parameters determined, \textit{e.g.} by a first-principles calculation,
we employ Rice-Mele to theoretically model two bands two bands near the Fermi energy.
The model parameters can be set to match the shape of the bands in a real material.
For example, the $w=-0.4$ valence (conduction) band depicted in Fig.~\ref{fig:init_w} 
has its minimum (maximum) at the $\Gamma$ point,
and under an energy scale $|v|=5$ meV a bandwidth of 4 meV,
as reported for the two flat bands closest to the Fermi energy in twisted bilayer graphene \cite{vishwanath}.

In Eq.~(\ref{eq:H}) we took $u=0.0$, so the Rice-Mele model reduces to the Su-Schriefer-Heeger model \cite{ssh}. The resulting energy bands are \cite{aachenthesis}
\begin{flalign}
    E_\pm (k) = \pm \sqrt{v^2 + w^2 + 2vw\cos(ka)} \,,
    \label{eq:dispersion_RiceMele}
\end{flalign}
with bandwidth $E_\text{band}=|w+\text{sgn}(w/v)v|-|w-\text{sgn}(w/v)v|$ and direct gap $E_{gap} = 2|w-\text{sgn}(w/v)v|$ between the $(+)$ and $(-)$ bands. For $\text{sgn}(w/v)=+1$ the gap is at the Brillouin zone boundary $k=\pm \pi/a$, while for $\text{sgn}(w/v)=-1$ the gap is at the $\Gamma$ point. 
Here we fix $v=-1$ and $\text{sgn}(w/v)=+1$ while detuning $w$. The detuned dispersion is shown in Fig.~\ref{fig:init_w}.
Compared to the trivial Rice-Mele limit $u = 0$, $v = w$,
\textit{i.e.} a gapless cosine band,
detuning $w$ allows us to mimic the flat-band limit of the band structure.

\section{Matrix Product States}
\label{app:MPS}

To construct a matrix product state (MPS), we work in the occupation number representation of a set of sites $l=1,2...L$, where the basis states are Slater determinants $|n_1...n_L\rangle$, with $n_l$ the occupation number of the $l^{th}$ site.
In this basis we can represent any many-body state $|\Phi\rangle$ using a rank-$L$ tensor $T^{n_1 ... n_L} $ such that \cite{chan}
\begin{flalign}
    |\Phi \rangle =& \sum_{n_1 ... n_L} T^{n_1 ... n_L} |n_1  ... n_L \rangle \,.
\end{flalign}
We can exactly transform a single rank-$L$ tensor into a product of $L$ rank-3 tensors according to 
\begin{flalign}
     T^{n_1 ... n_L}  =& \sum_{\alpha_1 ... \alpha_L}^{\chi_e} A^{(1)n_1}_{\alpha_L \alpha_1} A^{(2)n_2}_{\alpha_1 \alpha_2} ... A^{(L-1)n_{L-1}}_{\alpha_{L-2} \alpha_{L-1}}A^{(L)n_L}_{\alpha_{L-1} \alpha_L}  \,,
    \label{eq:mps_exact}
\end{flalign}
where in addition to the occupation number indices $n_l$ we have introduced virtual indices $\alpha_l$, all of which have dimensionality $\chi_e$. 
The transformation (\ref{eq:mps_exact}) is exact \cite{chan_inpractice} when
$\chi_e =4^{L/2}$ 
\footnote{
This holds for molecular orbitals where $n_l \in \{ 0, \uparrow_l, \downarrow_l, 2_l \}$ \cite{chan} while for spin orbitals or qubit systems $n_l \in \{0,1\}$ and $\chi_e = 2^{L/2}$.
}
, so the complexity of $|\Phi\rangle$ scales exponentially with system size. However, truncating the dimensionality of the virtual indices to the so-called bond dimension $\chi_b \ll \chi_e $ makes the product tractable \cite{feiguin}.
We are left with the
nonlinear
variational
MPS
\footnote{
    Following the convention, we dropped the contraction over $\alpha_L$ between $\textbf{A}^{(1)n_1}$ and $\textbf{A}^{(L)n_L}$.
}
\begin{flalign}
    |\Phi_\text{MPS}\rangle =& 
    \sum_{n_1 ... n_L}
\sum_{\alpha_1 ... \alpha_L}^{\chi_b} A^{(1)n_1}_{\alpha_1} A^{(2)n_2}_{\alpha_1 \alpha_2} 
...  A^{(L)n_L}_{\alpha_L-1} 
|n_1  ... n_L \rangle  \,.
    \label{eq:mps}
\end{flalign}
$|\Phi_\text{MPS}\rangle$ possesses only $\mathcal{O}(\chi_b^2 L)$ matrix elements \cite{chan, chan_inpractice}, as opposed to the exponentially scaling $\mathcal{O}(4^L)$ matrix elements in the exact $|\Phi\rangle$, but by harnessing the principle of locality it achieves controlled accuracy with increasing $\chi_b$ \cite{chan, feiguin}.

The fundamental objects of $|\Phi_\text{MPS}\rangle$ are the matrices
$\textbf{A}^{(l)n_l}$
whose matrix elements, indexed by $\alpha_l$, are the variational parameters \cite{chan, chan_inpractice} encoding physical correlations between different sites. $|\Phi_\text{MPS}\rangle$ can encode more correlations between sites whose corresponding matrices contract over a shared $\alpha_l$, so assigning an order $l=1,2...L$ to the sites impacts the results.
In other words, the MPS ansatz assumes a 1D structure to the physical correlation and imposes this through the site ordering. This makes site ordering straightforward for 1D or nearly 1D systems (\textit{e.g.} nanowires) and explains why td-DMRG is best applied to such systems \cite{chan_inpractice}.
For non-1D systems ordering the sites to optimize accuracy is challenging \cite{chan, chan_inpractice}. In our case, the sites discretizing the nanowire are already 1D.
%
%
To handle the MSQ degrees of freedom without sacrificing our 1D geometry,
we take each site in the scattering region unit cells $N_L \leq j \leq N_L+N_{FM}-1$, \textit{i.e.} all the sites exchange coupled to the MSQs by Eq.~(\ref{eq:H}),
and expand their local Hilbert space from purely fermionic
\begin{flalign}
    \{ |0\rangle_{j\mu}, |\uparrow\rangle_{j\mu}, |\downarrow\rangle_{j\mu}, |\uparrow\downarrow\rangle_{j\mu} \} \notag
\end{flalign}
to encompass the MSQ degrees of freedom:
\begin{flalign}
    \{ |0\rangle_{j\mu}, |\uparrow\rangle_{j\mu}, |\downarrow\rangle_{j\mu}, |\uparrow\downarrow\rangle_{j\mu} \} \otimes \{ |\uparrow\rangle_d, |\downarrow\rangle_d \} \,, \label{eq:supersiting}
\end{flalign}
a process known as `supersiting.'

\section{Time Evolution of Finite Quantum Systems}
\label{sec:finite}

\begin{figure}[t]
    \centering
    \includegraphics[width =1\linewidth]{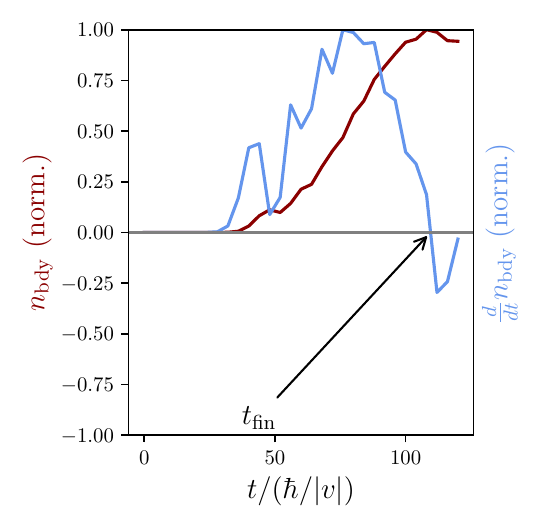}
    \caption[Onset of finite-size effects]{Onset of finite-size effects as the unphysical leftward reflection causes the occupation of the right boundary unit cell $n_\text{bdy}$ to plateau and its derivative $\tfrac{d}{dt} n_\text{bdy}$ to change sign. Curves show $w=-0.40$, $N_e=10$ simulation for qubits absent.}
    \label{fig:tfin}
\end{figure}

While the td-DMRG method unlocks access to more complicated systems with more itinerant electrons, we are still simulating finite systems and must consider finite size effects. This is a notable departure from the methods of Ciccarello \cite{ciccarello,  menezes}, who describe infinite leads at the cost of handling only one itinerant electron. 
In particular, our system contains $N_\text{total}$ sites, with the itinerant electrons starting in the leftmost of these and moving rightwards over time. 
The onset of finite size effects 
comes when the nanowire's right boundary at $j=N_\text{total} -1$ backscatters the electrons \cite{feiguin}, at the time we label $t_\text{fin}$.
Prior to $t_\text{fin}$, the transport of electrons across the scattering region faithfully describes the dynamics of electrons in an open nanowire \cite{feiguin} and can approximate steady-state transport quantities \cite{feiguin_conductance}. 

We quantify $t_\text{fin}$ using the time-dependent occupation of the right boundary unit cell,
\begin{flalign}
    n_\text{bdy}(t) = \sum_\mu \sum_\sigma \langle c_{j\mu\sigma}^\dagger c_{j\mu\sigma} \rangle_{j=N_\text{total}-1} \,.
\end{flalign}
Initially, this occupation monotonically increases due to the (physically normal) rightward electron flow. However, backscattering turns on (unphysical, caused by finite-size) leftward reflction that will eventually cancel out the rightward flow. This marks the presence of finite-size effects and allows us to define $t_\text{fin}$ as
\begin{flalign}
    \frac{d}{dt} n_\text{bdy}(t_\text{fin}) = 0 
    \,. \label{eq:tfin}
\end{flalign}
Fig.~\ref{fig:tfin} shows the determination of $t_\text{fin}$ for the $w=-0.40$ simulations. Note that we assign $t_\text{fin}$ to the discrete $t$ value closest to (\ref{eq:tfin}).

\section{Single-Particle Eigenstates}
\label{app:transport}

\begin{figure}[t]
    \centering
    \includegraphics[width =1\linewidth]{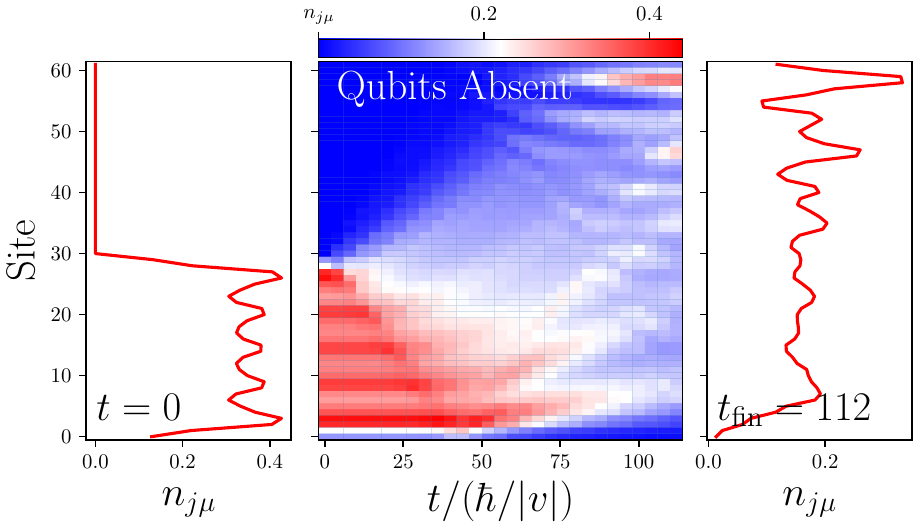}
    \caption[Site occupation]{(left) site occupation $n_{j\mu}$ preceding quantum quench, (middle) full time and space dynamics of $n_{j\mu}$ following quantum quench, and (right) $n_{j\mu}$ at the end of the simulation.
    Throughout, Hamiltonian parameters are $w=-0.40$ and $N_e=10$, geometric parameters are $N_L=N_R=15$, $N_\text{total} = 31$ and td-DMRG parameters are $\chi_b = 250$, $dt=0.1$.}
    \label{fig:occ_profile}
\end{figure}

For $t<0$ Hamiltonian contains
no interaction between the MSQs and itinerant electrons,
so $\hat{H}(t<0)$ can be broken into a part acting on MSQs [the last line of Eq.~(\ref{eq:H0})] and a part acting on itinerant electrons [the $v$ and $w$ terms in Eq.~(\ref{eq:H}) and the first line of Eq.~(\ref{eq:H0})].
The electronic part of $\hat{H}(t<0)$ commutes with the electron momentum operator as well as the electron spin in the $z$-direction 
$\sum_{j\mu}(c_{j\mu\uparrow}^\dagger c_{j\mu\uparrow} - c_{j\mu\downarrow}^\dagger c_{j\mu\downarrow})/2$, 
so we label the $m^{th}$ $t<0$ single-particle eigenstate $|k_m \sigma_e^m \rangle$,
where $\hbar k_m$ is the momentum of the state and $\sigma_e^m = \pm\hbar/2$ is the projection of its spin along the $z$-axis.
Analogously to the Bloch states propagating in the crystal lattice of a metal or semiconductor,
the single-particle eigenstates $|k_m \sigma_e^m \rangle $ are the modes through which the itinerant electrons propagate along the nanowire.
The energy eigenvalues $E$ of these eigenstates are discrete due to the finite size of the system, but still satisfy the Rice-Mele dispersion [Eq.~(\ref{eq:dispersion_RiceMele})] from which follows the density of states
\begin{flalign}
    \rho(E) = 2/\pi| dE/dk_m| \,.
\end{flalign} 
Under Eq.~(\ref{eq:H}) the energies are independent of $\sigma_e^m$ and so twofold degenerate.
The occupation function for a spin-unpolarized system ($N_e$ even) is
\begin{flalign}
    n_m (E) = \left\{ 
    \begin{matrix}
        2 & m \leq N_e \\
        0 & m > N_e
    \end{matrix}
    \right\}
\end{flalign}
and the Fermi energy is $E_F =E_{N_e}$.
From the single-particle eigenstates we can construct the
$t<0$ many-body state
\begin{flalign}
    |\Psi(t<0)\rangle = |n_1 ... n_m ...  \rangle \otimes 
    |\psi_\text{qubit}\rangle
    \label{eq:t0manybody}
\end{flalign}
where $|n_1 ... n_m ... \rangle$ is a Slater determinant in the occupation number representation of the single-particle eigenstates
and $|\psi_\text{qubit}\rangle$ is a maximally entangled state of the form (\ref{eq:cicc_state_alt}).
Note that $\hat{H}_0$ dictates the phase of $|\psi_\text{qubit}\rangle$ according to 
\begin{align}
    \phi_\text{ent} = \cos^{-1}\left(J_\text{ent}/\sqrt{J_\text{ent}^2 + J_\text{ent}'^2}
    \right)
    \label{eq:define_phient}
\end{align}
and that $\phi_\text{ent}$ can be read from the observable 
\begin{flalign}
    \langle \psi_\text{qubit}| (\textbf{S}_1 + \textbf{S}_2)^2 |\psi_\text{qubit} \rangle = 1+\cos(\phi_\text{ent})\,.
\end{flalign}

The $t<0$ many-body state $|\Psi(t<0)\rangle$ is a key variable in our simulations, determining the energies and momenta of the electrons during the transport.
We depict $|\Psi(t<0)\rangle$ in real space in Fig.~\ref{fig:occ_profile} and in energy space in Fig.~\ref{fig:init_w}.
In the latter, we observe how detuning $w$ changes the electronic properties, gapping the single-particle eigenstates and narrowing the bandwidth to $E_\text{band}=0.8$. 
In general we expect the quantum spin valve efficiency to vary with these electronic properties \cite{ciccarello} 
.
The full distribution of single-particle energies and momenta can be seen in Fig. \ref{fig:init_w}.
A simple, physical way to describe the single-particle eigenstates as we change $|\Psi(t<0)\rangle$ 
by detuning $w$
is the density of states at the Fermi energy $\rho(E_F)$, which tells us about the propagation of the highest energy itinerant electron. 
The momentum of a given electron is inversely proportional to its density of states, and when $\rho(E_F)$ is small we expect faster propagating electrons.
For fixed $N_e$, $\rho(E_F)$ has a one-to-one correspondence with $w$ as seen the inset of Fig.~\ref{fig:efficiency} where decreasing $w$ increases $\rho(E_F)$ and flattens the band.
The Rice-Mele hopping $w$ therefore serves as a knob in the Hamiltonian that we can use to engineer desirable electronic properties, such as enhanced $\rho(E_F)$.

\section{Code Availability}

Full td-DMRG code used in all simulations available upon request.

\end{suppinfo}

\bibliography{referenceFile}

\end{document}